\documentclass[aps,showpacs,superscriptaddress,
preprintnumbers,10pt,onecolumn]{revtex4}%
\usepackage{amsfonts}
\usepackage{amsmath}
\usepackage{hyperref}
\usepackage{scalefnt}
\usepackage{amssymb}
\usepackage{graphicx}%
\newtheorem{theorem}{Theorem}

\newtheorem{lemma}[theorem]{Lemma}

\newenvironment{proof}[1][Proof]{\noindent\textbf{#1.} }{\ \rule{0.5em}{0.5em}}

\begin{document}
\title{Universal quantum computation with unlabeled qubits}
\author{Simone Severini}
\email{ss54@york.ac.uk}
\affiliation{Department of Mathematics and Department of Computer Science, University of
York, Heslington, YO10 5DD York, U.K.}

\begin{abstract}
We show that an $n$-th root of the Walsh-Hadamard transform (obtained from the
Hadamard gate and a cyclic permutation of the qubits), together with two
diagonal matrices, namely a local qubit-flip (for a fixed but arbitrary qubit)
and a non-local phase-flip (for a fixed but arbitrary coefficient), can do
universal quantum computation on $n$ qubits. A quantum computation, making use
of $n$ qubits and based on these operations, is then a word of variable
length, but whose letters are always taken from an alphabet of cardinality
three. Therefore, in contrast with other universal sets, no choice of qubit
lines is needed for the application of the operations described here. A
quantum algorithm based on this set can be interpreted as a discrete diffusion
of a quantum particle on a de Bruijn graph, \emph{corrected on-the-fly} by
auxiliary modifications of the phases associated to the arcs.

\end{abstract}
\maketitle

\section{Introduction}

The study of universality in quantum computation goes back to \cite{d}. In the
circuit model, universality has been considered by a number of papers (see
\cite{ah, shi} and the references therein). Probably, the simplest universal
set of gates consists of the Hadamard gate $H$ together with the Toffoli gate
$T$ \cite{shi}. Of course, in order to have universality, we need the freedom
of applying $H$ and $T$ to arbitrary qubits of the computer: $H$ to any qubit
and $T$ to any three qubits. In fact, out of measurement processes, every
computational step is induced by a unitary, obtained by tensoring together
$H$'s, $T$'s and identity matrices. The number of different unitaries obtained
in this way is then a function of the number of qubits.

\bigskip

In this paper, we define a universal set of unitary quantum operations
depending on the computational space of the machine. This means that the
unitaries change whenever the total number of qubits changes. Since the
operations act globally on all qubits, they do not require the choice qubit
lines at each computational step. This is in contrast to standard finite
universal sets (\emph{e.g.}, $H$ and $T$) where the gates reman fixed but may
be applied to varying choices of qubit lines. The set defined here is composed
by an $n$-th root of the Walsh-Hadamard transform (constructed from the
Hadamard gate and a cyclic permutation of the qubits), and two diagonal
matrices:\ a (local) qubit-flip (for a fixed but arbitrary qubit) and a
(non-local) phase-flip (for a fixed but arbitrary coefficient). To prove that
the defined set is universal on $n$ qubits, we reduce it to $H$ and $T$,
acting on any qubit and any three qubits. The characteristic property of the
set is that the choice of the qubits to which apply $H$ and $T$ is not
reflected into the structures of the unitaries. This observation would like to
be a justification to the title of the paper. Of course, the set described
here is inconvenient from the physical impementation point of view, because it
requires nonlocal interaction between qubits (which is physically expensive).
However, it may be useful to remark that the set provides a mathematical
framework, in which doing universal quantum computation is conctructing words
whose letters are unitaries (two of which commute) taken from an alphabet of
cardinality three. So, the outcome of a computation depends on the length of
the word and the order of the letters. This has some flavour that reminds of
quantum finite state automata and other sequential machines (see \emph{e.g.}
\cite{mc, gu}).

\bigskip

The remainder of this paper is organized as follows. In Section 2, we give
some preliminary definitions. In Section 3, we formally state and prove the
main result. This is done by proving universality with a reduction to
$\{H,T\}$. It is not difficult to verify that the $n$-th root of the
Walsh-Hadamard transform respects the topology of the de Bruijn graph. Namely,
the $ij$-th entry of this unitary is nonzero if and only if there is a
directed edge from the vertex labeled $i$ to the vertex labeled $j$, in the de
Bruijn graph on $2^{n}$ vertices. In Section 4, in virtue of this observation,
we point out that any quantum algorithm can be seen as the discrete diffusion
of a quantum particle on a de Bruijn graph, \emph{corrected on-the-fly} by a
qubit-flip and a phase-flip, both fixed but arbitrary. This reminds of the
context of discrete quantum walks \cite{am} or the processes studied in
\cite{ko}. A natural open question would be to prove that the $n$-th root of
the Walsh-Hadamard transform and a phase-flip (for a fixed but arbitrary
coefficient) form a universal set.

\section{Definitions}

In this section we introduce some preliminary definitions.

\bigskip

\noindent\textbf{Definition of }$V_{n}$\textbf{.} We denote by $V_{n}$ a
square matrix of dimension $2^{n}$ such that $[V_{n}]_{i,j}\in\{0,\pm\frac
{1}{\sqrt{2}}\}$ and with exactly the following nonzero entries:%
\begin{equation}%
\begin{tabular}
[c]{l}%
$\lbrack V_{n}]_{1,1}=[V_{n}]_{1,2^{n-1}+1}=[V_{n}]_{2,1}=\frac{1}{\sqrt{2}}%
,$\\
$\lbrack V_{n}]_{2,2^{n-1}+1}=-\frac{1}{\sqrt{2}},$\\
$\lbrack V_{n}]_{3,2}=[V_{n}]_{3,2^{n-1}+2}=[V_{n}]_{4,2}=\frac{1}{\sqrt{2}}%
,$\\
$\lbrack V_{n}]_{4,2^{n-1}+2}=-\frac{1}{\sqrt{2}},$\\
$\vdots$\\
$\lbrack V_{n}]_{2^{n}-1,2^{n-1}}=[V_{n}]_{2^{n}-1,2^{n}}=[V_{n}%
]_{2^{n},2^{n-1}}=\frac{1}{\sqrt{2}},$\\
$\lbrack V_{n}]_{2^{n},2^{n}}=-\frac{1}{\sqrt{2}}.$%
\end{tabular}
\ \ \ \ \ \ \ \ \ \label{defw}%
\end{equation}
The matrix $V_{n}$ is real-orthogonal and it is an $n$-th root of
$H_{n}:=H^{\otimes n}$, where $H$ is the 1-qubit hadamard gate:
\begin{equation}
H:=\frac{1}{\sqrt{2}}\left(
\begin{array}
[c]{cc}%
1 & 1\\
1 & -1
\end{array}
\right)  .
\end{equation}
For example,
\begin{equation}%
\begin{tabular}
[c]{l}%
$V_{2}=\frac{1}{\sqrt{2}}\left(
\begin{array}
[c]{rrrr}%
1 & 0 & 1 & 0\\
1 & 0 & -1 & 0\\
0 & 1 & 0 & 1\\
0 & 1 & 0 & -1
\end{array}
\right)  $%
\end{tabular}
\ \ \ \ \ \ \ \ \ \ \ \ \ \
\end{equation}
and $V_{2}^{2}=H$. An alternative and more direct definition of $V_{n}$ can be
given as follows. Let $S_{n}$ be the full symmetric group on the set
$\{1,2,...,n\}$. We denote permutations of length $n$ as ordered sets. For
example, the elements of $S_{3}$ are $\left(  1,2,3\right)  $, $\left(
1,3,2\right)  $, $\left(  3,2,1\right)  $, $\left(  2,3,1\right)  $, $\left(
2,1,3\right)  $, and $(3,1,2)$. With an abuse of our notation, their regular
permutation representations are denoted in the same way. The matrix $V_{n}$ is
defined as%
\begin{equation}
V_{n}:=P\cdot\left(  H\otimes I^{\otimes n-1}\right)  , \label{defm}%
\end{equation}
where%
\begin{equation}
P=(1,3,...,2^{n}-1,2,4,...,2^{n}).
\end{equation}
This permutation is nothing but a cyclic-shift of the qubits:
\begin{equation}
P:|a_{1}a_{2}...a_{n}\rangle\longrightarrow|a_{n-1}a_{1}...a_{n-2}\rangle.
\end{equation}
This explains why $V_{n}^{n}=H_{n}$ ($H_{n}:=H^{\otimes n}$). It is easy to
verify that Eq. \ref{defw} and Eq. \ref{defm} define the same matrices. Notice
that, when $n=2$, the permutation $P$ is the Swap-gate:
\begin{equation}
(1,3,2,4)=\left(
\begin{array}
[c]{rrrr}%
1 & 0 & 0 & 0\\
0 & 0 & 1 & 0\\
0 & 1 & 0 & 0\\
0 & 0 & 0 & 1
\end{array}
\right)  .
\end{equation}

\bigskip

\noindent\textbf{Definition of }$P_{n}(k)$\textbf{.} Let us denote by
$P_{n}(k)$ the $2^{n}\times2^{n}$ matrix defined by%
\begin{equation}
\lbrack P_{n}(k)]_{i,j}:=\left\{
\begin{tabular}
[c]{rr}%
$0$ & if $i\neq j$;\\
$-1$ & if $i=k$ with $k\in\{1,...,2^{n}\}$;\\
$1$ & otherwise.
\end{tabular}
\ \ \ \ \ \ \ \ \ \ \ \ \ \ \right.
\end{equation}
The matrix $P_{n}(k)$ is the Pauli operator $Z=\left(
\begin{array}
[c]{cc}%
1 & 0\\
0 & -1
\end{array}
\right)  $ on the $n$-th qubit controlled by all other qubits $1,2,...,n-1$
and then conjugated by the permutation that interchanges dimensions $k$ and
$2^{n}$. For example,
\begin{equation}
P_{2}(3)=\left(
\begin{array}
[c]{cccc}%
1 & 0 & 0 & 0\\
0 & 1 & 0 & 0\\
0 & 0 & -1 & 0\\
0 & 0 & 0 & 1
\end{array}
\right)  =I_{2}\oplus XZX,
\end{equation}
where $X$ is the Pauli operator $X=\left(
\begin{array}
[c]{cc}%
0 & 1\\
1 & 0
\end{array}
\right)  $.

\bigskip

\noindent\textbf{Definition of }$F_{n}(k)$\textbf{. }Let us denote by
$F_{n}(k)$ the $2^{n}\times2^{n}$ matrix defined by%
\begin{equation}
\lbrack F_{n}(k)]_{i,j}:=\left\{
\begin{tabular}
[c]{rr}%
$0$ & if $i\neq j$;\\
$-1$ & if the $k$-th qubit is $1$;\\
$1$ & otherwise.
\end{tabular}
\ \ \ \ \ \ \ \ \ \ \ \right.
\end{equation}
The matrix $F_{n}(k)$ is the Pauli operator $Z$ acting on the $k$-th qubit and
the identity on all other qubits. For example,
\begin{equation}%
\begin{tabular}
[c]{lll}%
$F_{2}(1)=\left(
\begin{array}
[c]{cccc}%
1 & 0 & 0 & 0\\
0 & -1 & 0 & 0\\
0 & 0 & 1 & 0\\
0 & 0 & 0 & -1
\end{array}
\right)  =I\otimes Z$ & and & $F_{2}(2)=\left(
\begin{array}
[c]{cccc}%
1 & 0 & 0 & 0\\
0 & 1 & 0 & 0\\
0 & 0 & -1 & 0\\
0 & 0 & 0 & -1
\end{array}
\right)  =Z\otimes I.$%
\end{tabular}
\ \ \ \ \ \ \ \ \ \ \ \ \ \
\end{equation}

\section{Main result}

In this section we prove the following theorem.

\begin{theorem}
\label{ma}The set
\begin{equation}
B=\{V_{n},P_{n}(i),F_{n}(j)\},
\end{equation}
for fixed but arbitrary $i\in\{1,...,2^{n}\}$ and $j\in\{1,...,n\}$, is
universal for quantum computation on $n$ qubits.
\end{theorem}

By this theorem, any quantum computation in the circuit model can be seen as a
word whose letters are taken from the alphabet $B$.

\bigskip

The \emph{Toffoli gate} $T$ is the $3$-qubit gate defined as
$T:(a,b,c)\longrightarrow(a,b,ab\oplus c)$, where $\oplus$ denotes addition
modulo $2$ and $a,b,c\in\{0,1\}$. The matrices defined by the Toffoli gate are
then a special kind of transposition. If we label the states of the
computational basis in lexicographic order, the matrices defined by the
Toffoli gate on $3$ qubits are%
\begingroup
\scalefont{0.8}%
\begin{equation}%
\begin{tabular}
[c]{llll}%
$\left(
\begin{array}
[c]{cccccccc}%
1 & 0 & 0 & 0 & 0 & 0 & 0 & 0\\
0 & 1 & 0 & 0 & 0 & 0 & 0 & 0\\
0 & 0 & 1 & 0 & 0 & 0 & 0 & 0\\
0 & 0 & 0 & 1 & 0 & 0 & 0 & 0\\
0 & 0 & 0 & 0 & 1 & 0 & 0 & 0\\
0 & 0 & 0 & 0 & 0 & 1 & 0 & 0\\
0 & 0 & 0 & 0 & 0 & 0 & 0 & 1\\
0 & 0 & 0 & 0 & 0 & 0 & 1 & 0
\end{array}
\right)  ,$ & $\left(
\begin{array}
[c]{cccccccc}%
1 & 0 & 0 & 0 & 0 & 0 & 0 & 0\\
0 & 1 & 0 & 0 & 0 & 0 & 0 & 0\\
0 & 0 & 1 & 0 & 0 & 0 & 0 & 0\\
0 & 0 & 0 & 0 & 0 & 0 & 0 & 1\\
0 & 0 & 0 & 0 & 1 & 0 & 0 & 0\\
0 & 0 & 0 & 0 & 0 & 1 & 0 & 0\\
0 & 0 & 0 & 0 & 0 & 0 & 1 & 0\\
0 & 0 & 0 & 1 & 0 & 0 & 0 & 0
\end{array}
\right)  $ & and & $\left(
\begin{array}
[c]{cccccccc}%
1 & 0 & 0 & 0 & 0 & 0 & 0 & 0\\
0 & 1 & 0 & 0 & 0 & 0 & 0 & 0\\
0 & 0 & 1 & 0 & 0 & 0 & 0 & 0\\
0 & 0 & 0 & 1 & 0 & 0 & 0 & 0\\
0 & 0 & 0 & 0 & 1 & 0 & 0 & 0\\
0 & 0 & 0 & 0 & 0 & 0 & 0 & 1\\
0 & 0 & 0 & 0 & 0 & 0 & 1 & 0\\
0 & 0 & 0 & 0 & 0 & 1 & 0 & 0
\end{array}
\right)  .$%
\end{tabular}
\end{equation}
%

\endgroup
The set $B^{\prime}=\{H,T\}$ is universal for quantum computation \cite{ah,
shi}. We prove Theorem \ref{ma} by showing that the set $B$ is equivalent to
the set $B^{\prime}$, once the number of qubits has been fixed. In order to
verify this equivalence, we show that any expression made by the tensor
product of $H$, $T$, and the $2\times2$ identity matrix $I_{2}$ corresponds to
a sequence of the three elements of $B$ defined in the statement of the above
theorem. We will consider the set
\begin{equation}
D_{n}:=\{P_{n}(i):1\leq i\leq2^{n}\}.
\end{equation}

\begin{lemma}
\label{se1}The set $A=V_{n}\cup D_{n}$ is universal for quantum computation on
$n$ qubits.
\end{lemma}

\begin{proof}
By the definition of $V_{n}$ (Eq. \ref{defm} above), we have
\begin{equation}
P^{-1}\cdot V_{n}=H\otimes I^{\otimes n-1}, \label{hadm}%
\end{equation}
where
\[
P^{-1}=(1,2^{n-1}+1,2,2^{n-1}+2,3,...,2^{n-1},2^{n})=(1,3,...,2^{n}%
-1,2,4,...,2^{n})^{-1}.
\]
This means that if we can construct the matrix $P^{-1}$ then we can also
construct the Hadamard gate. Firstly, observe that
\[
V_{n}^{-1}=V_{n}^{2n-1}=V_{n}^{T},
\]
since
\[
V_{n}^{2n}=V_{n}^{n}V_{n}^{n}=H_{n}H_{n}=I^{\otimes n}.
\]
The matrix $V_{n}^{T}$ can be then obtained directly from $V_{n}$. The nonzero
entries in the last two rows of $V_{n}$ are
\[%
\begin{tabular}
[c]{l}%
$\lbrack V_{n}]_{2^{n}-1,2^{n-1}}=[V_{n}]_{2^{n}-1,2^{n}}=[V_{n}%
]_{2^{n},2^{n-1}}=\frac{1}{\sqrt{2}},$\\
$\lbrack V_{n}]_{2^{n},2^{n}}=-\frac{1}{\sqrt{2}}.$%
\end{tabular}
\ \ \
\]
The nonzero entries in the last two columns of $Q_{n}=P_{n}(2^{n})\cdot
V_{n}^{T}$ are
\[%
\begin{tabular}
[c]{l}%
$\lbrack Q_{n}]_{2^{n-1},2^{n}-1}=[Q_{n}]_{2^{n},2^{n}-1}=[P_{n}(2^{n})\cdot
V_{n}^{T}]_{2^{n-1},2^{n}}=\frac{1}{\sqrt{2}},$\\
$\lbrack Q_{n}]_{2^{n},2^{n}}=-\frac{1}{\sqrt{2}}.$%
\end{tabular}
\ \ \
\]
Since $H\cdot\frac{1}{\sqrt{2}}\left(
\begin{array}
[c]{cc}%
1 & 1\\
-1 & 1
\end{array}
\right)  =X$, it follows that%
\begin{equation}
V_{n}Q_{n}=V_{n}P_{n}(2^{n})V_{n}^{T}=I_{2^{n}-2}\oplus X=(1,2,...,2^{n}%
-2,2^{n},2^{n}-1), \label{to}%
\end{equation}
which is indeed the Toffoli gate. Now, given that%
\[
S_{2^{n}}=\langle(1,2,...,2^{n}-2,2^{n},2^{n}-1),(2,3,...,2^{n},1)\rangle,
\]
if we can construct the cyclic permutation $(2,3,...,2^{n},1)$ then we will
have the full symmetric group $S_{2^{n}}$. If we can construct $S_{2^{n}}$
then we will get $P^{-1}$ and all the permutations that we need for applying
$H$ and $T$ to arbitrary qubits. The permutation $(2,3,...,2^{n},1)$ can be
constructed with the following procedure (most probably not optimal):

\begin{enumerate}
\item Let $M_{1}=H_{n}F_{n}(n)H_{n}=X\otimes I^{\otimes n-1}$. By applying
elements from $D_{n}$, tranform the bottom-left block of $M_{1}$ into the
diagonal matrix $dia(p_{1},...,p_{2^{n-3}},-p_{2^{n-3}+1},...,-p_{2^{n-2}})$,
where $p_{i}=(1,-1)$. Let $M_{2}^{\prime}$ be the matrix obtained in this way.

\item Let $M_{2}=V_{n}^{-1}M_{2}^{\prime}V_{n}$. By applying elements from
$D_{n}$, transform the top-right block of $M_{2}$ into the identity matrix and
the bottom-left block into the diagonal matrix $dia(p_{1},...,p_{2^{n-4}%
},-p_{2^{n-4}+1},...,-p_{2^{n-3}})$.

\item Repeating the second step $n-2$ times one gets the permutation matrix
$(2,3,...,2^{n},1)$. (This can be easily checked with any computer algebra system.)
\end{enumerate}

An example with three qubits may help to clarify the procedure:
\begin{align*}
H_{3}\left(  1,2,3,4,\overline{5},\overline{6},\overline{7},\overline
{8}\right)  H_{3}  &  =\left(  5,6,7,8,1,2,3,4\right)  ,\\
V_{3}^{-1}\left(  5,6,7,8,1,\overline{2},\overline{3},4\right)  V_{3}  &
=\left(  3,4,5,\overline{6},7,8,1,\overline{2}\right)  ,\\
V_{3}^{-1}\left(  3,4,5,6,7,8,1,\overline{2}\right)  V_{3}  &  =\left(
2,3,4,5,6,7,8,1\right)  .
\end{align*}
The notation is easily explained:%
\[
\left(  5,6,7,8,1,\overline{2},\overline{3},4\right)  =\left(
\begin{tabular}
[c]{c|c}%
$\mathbf{0}$ & $%
\begin{array}
[c]{cccc}%
1 & 0 & 0 & 0\\
0 & 1 & 0 & 0\\
0 & 0 & 1 & 0\\
0 & 0 & 0 & 1
\end{array}
$\\\hline
\multicolumn{1}{r|}{$%
\begin{array}
[c]{cccc}%
1 & 0 & 0 & 0\\
0 & -1 & 0 & 0\\
0 & 0 & -1 & 0\\
0 & 0 & 0 & 1
\end{array}
$} & $\mathbf{0}$%
\end{tabular}
\ \ \right)  .
\]
By the above constructions, we have the following fact: the matrix group
$G=\langle V_{n},D_{n}\rangle$ has the regular permutation representation of
$S_{2^{n}}$ as a subgroup. Then $T$, obtained with Eq. \ref{to}, can be
applied to any three qubits; $H$, obtained from Eq. \ref{hadm}, can be applied
to to any qubit. Since $B^{\prime}=\{H,T\}$ is universal, the lemma follows.
\end{proof}

\begin{lemma}
\label{se}The matrix group $G=\langle V_{n},P_{n}(i),F_{n}(j)\rangle$, for
fixed but arbitrary $i\in\{1,...,2^{n}\}$ and $j\in\{1,...,n\}$, contains the
set $D_{n}$.
\end{lemma}

\begin{proof}
By the definitions,%

\begin{align*}
V_{n}^{k}F_{n}(n)V_{n}^{-k}  &  =\left(  P\left(  H\otimes I^{\otimes
n-1}\right)  \right)  ^{k}\cdot\left(  Z\otimes I^{\otimes n-1}\right)
\cdot\left(  P\left(  H\otimes I^{\otimes n-1}\right)  \right)  ^{-k}\\
&  =P^{k}\left(  H\otimes I^{\otimes n-1}\right)  \cdot\left(  Z\otimes
I^{\otimes n-1}\right)  \cdot\left(  H\otimes I^{\otimes n-1}\right)  P^{-k}\\
&  =P^{k}\left(  X\otimes I^{\otimes n-1}\right)  P^{-k}\\
&  =I\otimes\cdots\otimes I\otimes X\otimes I\cdots\otimes I,
\end{align*}
where $X$ is at the $k$-th position of the tensor product. For example,
$V_{3}^{1}F_{3}(3)V_{3}^{-1}=I\otimes I\otimes X$. Let $H$ be the matrix group
generated by the matrices $I\otimes\cdots\otimes I\otimes X,I\otimes
\cdots\otimes I\otimes X\otimes I,X\otimes I\otimes\cdots\otimes I$. The group
$H$ is isomorphic to $\mathbb{Z}_{2}^{n}$ (indeed, the above matrices are the
permutation representations of the standard generators of $\mathbb{Z}_{2}^{n}%
$). Since, for every $i,j\in\{1,...,2^{n}\}$, there is an element of $H$
sending $P_{n}(i)$ to $P_{n}(j)$, we can construct $D_{n}$, that is the set of
all $2^{n}\times2^{n}$ diagonal matrices with entries $1$ and $-1$. We
considered $F_{n}(n)$, but we could also take $F_{n}(j)$, for any
$j\in\{1,...,n\}$, without loss of generality.
\end{proof}

\bigskip

Theorem \ref{ma} is a consequence of Lemma \ref{se1} together with Lemma
\ref{se}.

\section{A relation with de Bruijn graphs}

Let $\Sigma$ be an alphabet of cardinality $d$ and let $\Sigma_{n}^{\ast}$ be
the set of all words of length $n$ over $\Sigma$. The $d$\emph{-ary }%
$n$\emph{-dimensional de Bruijn} \emph{graph} is a directed graph denoted by
$B(d,n)$ and defined as follows \cite{h}: the set of vertices is $\Sigma
_{k}^{\ast}$; there is an arc from $i$ to $j$ if and only if the last $n-1$
letters of $i$ are the same as the first $n-1$ letters of $j$. The graph
$B(2,n)$ is called (\emph{directed})\ \emph{binary de Bruijn graph}. Notice
that $B(2,n)$ has exactly two loops: one loop is at the vertex $0\ldots0$; the
other one at the vertex $1\ldots1$. These graphs have important applications
in cryptography and distributed computing. In particular, they provide some of
the best-known topologies for communication networks. For example, the Galileo
space probe of NASA used a network based on a de Bruijn graph to implement a
signal decoder \cite{m}. Let $M_{n}$ be the adjacency matrix of $B(2,n)$. The
rows and the columns of this matrix can be ordered in such a way that%
\[%
\begin{tabular}
[c]{l}%
$\lbrack M_{n}]_{1,1}=[M_{n}]_{1,2^{n-1}+1}=[M_{n}]_{2,1}=[M_{n}%
]_{2,2^{n-1}+1}=1,$\\
$\lbrack M_{n}]_{3,2}=[M_{n}]_{3,2^{n-1}+2}=[M_{n}]_{4,2}=[M_{n}%
]_{4,2^{n-1}+2}=1,$\\
$\vdots$\\
$\lbrack M_{n}]_{2^{n}-1,\frac{1}{2}2^{n}-1}=[M_{n}]_{2^{n}-1,2^{n}}%
=[M_{n}]_{2^{n},\frac{1}{2}2^{n}-1}=[M_{n}]_{2^{n},2^{n}}=1.$%
\end{tabular}
\
\]
Tanner \cite{t} pointed out that the matrix $V_{n}$ is obtained from the
matrix $M_{n}$ by negating the entries $[M_{n}]_{2,2^{n-1}+1},[M_{n}%
]_{4,2^{n-1}+2},...,[M_{n}]_{2^{n},2^{n}}$ and rescaling $M_{n}$ by $\frac
{1}{\sqrt{2}}$. It is easy to see that a simple random walk on $B(d,n)$
converges very quickly to uniformity, in fact it is perfectly mixed after $n$
steps. \emph{Simple}\ means that at each vertex the walker chooses to cross an
incident edge by tossing a fair die with $d$ faces. We associate the vertices
of $B(2,n)$ with the elements of the computational basis $|0\rangle
\equiv|0...0\rangle,|1\rangle\equiv|0...01\rangle,...,|2^{n-1}\rangle
\equiv|1...1\rangle$. For every $|i\rangle$ and $|j\rangle$, there is a
diagonal matrix $Q\in D_{n}$ such that $H_{n}QH_{n}|i\rangle=V_{n}^{n}%
QV_{n}^{n}|i\rangle=|j\rangle$. We may interpret this process as a discrete
quantum walk on $B(2,n)$ induced by $V_{n}$ and \emph{corrected on-the-fly} by
an appropriate diagonal unitary with $\pm1$ entries. If the walk is not
corrected then $V_{n}^{2n}|i\rangle=|i\rangle$ for every $i$, given that
$H_{n}$ is symmetric. The walk on $B(2,n)$ is perfectly mixed at the $n$-th
step, but it can be driven with probability $1$ to any vertex in exactly $2n$
steps. For example, the following is an algorithm that takes the state
$|0\rangle$ to the state $|2^{n}-1\rangle$ in $2n$ steps:
\[%
\begin{tabular}
[c]{l}%
$V_{n}^{n}|0\rangle=|+\rangle^{\otimes n};$\\
$F_{1}(1)^{\otimes n}|+\rangle^{\otimes n}=|\psi\rangle;$\\
$V_{n}^{n}|\psi\rangle=|2^{n}-1\rangle.$%
\end{tabular}
\
\]
(A curiosity: the positions of the minus sign in the state $|\psi\rangle$
correspond to the numbers with an odd number of $1$'s in their binary
expansion (A007413 \cite{s})). A generalization for any $|i\rangle$ and
$|j\rangle$ is straightforward.

\bigskip

\emph{Aknowledgments. }I\ would like to thank the anonymous referees. Their
comments helped me to improve the presentation of this paper.


\begin{thebibliography}{99}                                                                                               %


\bibitem {ah}D. Aharonov, Simple Proof that Toffoli and Hadamard are Quantum
Universal, quant-ph/0301040.

\bibitem {am}A. Ambainis, Quantum walks and their algorithmic applications, quant-ph/0403120.

\bibitem {d}D. Deutsch, Quantum computational networks, \emph{Proc. Roy. Soc.
Lond.} A \textbf{425} 73-90, 1989.

\bibitem {gu}S. Gudder, Quantum automata: an overview,\emph{ Internat. J.
Theoret. Phys.} \textbf{38} (1999), no. 9, 2261--2282.

\bibitem {h}M.-C. Heydemann, Cayley graphs and interconnection networks,
\emph{Graph symmetry (Montreal, PQ, 1996)}, 167--224, NATO Adv. Sci. Inst.
Ser. C Math. Phys. Sci., 497, \emph{Kluwer Acad. Publ., Dordrecht}, 1997.

\bibitem {ko}J. Ko\v{s}\'{\i}k, Scattering Quantum Random Walk, \emph{Optics
and Spectroscopy }\textbf{99}, 224-226 (2005).

\bibitem {mc}C. Moore, J. P. Crutchfield, Quantum automata and quantum
grammars, \emph{Theoret. Comput. Sci.} \textbf{237} (2000), no. 1-2, 275--306.

\bibitem {m}B. Mukherjee, \emph{Optical Communication Networks}, Series on
Computer Communications, McGraw-Hill, New York, 1997.

\bibitem {shi}Y. Shi, Both Toffoli and controlled-Not need little help to do
universal quantum computation, quant-ph/0205115.

\bibitem {s}N. J. A. Sloane, (2005), The On-line Encyclopedia of Integer
Sequences, www.research.att.com/\symbol{126}njas/sequences/.

\bibitem {t}G. Tanner, Spectral statistics for unitary transfer matrices of
binary graphs, \emph{J. Phys. A} \textbf{33} (2000), no. 18, 3567--3585.
\end{thebibliography}
\end{document}